\documentclass[namedreferences]{solarphysics}
\usepackage[optionalrh,natbib]{spr-sola-addons} % For Solar Physics 
\usepackage{graphicx}        % For eps figures, newer & more powerfull
\usepackage{color}           % For color text: \color command
\usepackage{url}             % For breaking URLs easily trough lines
\usepackage[normalem]{ulem}
\newcommand{\aap}{    {\it Astron. Astrophys.}}

\newcommand{\apj}{    {\it Astrophys. J.}}
\newcommand{\apjl}{   {\it Astrophys. J. Lett.}}

\newcommand{\mnras}{  {\it Mon. Not. Roy. Astron. Soc.}}

\newcommand{\solphys}{{\it Solar Phys.}}

\newcommand{\ssr}{    {\it Space Sci. Rev.}}

\begin{document}

\begin{article}

\begin{opening}

\title{The Sub-Surface Structure of a Large Sample of Active Regions}

\author{C.S.~\surname{Baldner}$^{1}$\sep
	R.S.~\surname{Bogart}$^{1}$\sep
	S.~\surname{Basu}$^{2}$}
\runningauthor{C.S.~Baldner {\it et al.\/}}
\runningtitle{Structure of Active Regions}

\institute{$^{1}$ Hansen Experimental Physics Laboratory, Stanford University, Stanford, CA, 94305-4085, USA, email: baldner@stanford.edu \\
	$^{2}$ Department of Astronomy, Yale University, P.O. Box 208101, New Haven, CT 06520-8101, USA  \\
	}

\begin{abstract}
We employ ring-diagram analysis to study the sub-surface thermal structure 
of active regions. We present results using a large number of active regions 
over the course of Solar Cycle 23. We present both traditional inversions of 
ring-diagram frequency differences, with a total sample size of 264, and a 
statistical study using Principal Component Analysis. We confirm earlier 
results on smaller samples that sound speed and adiabatic index are changed 
below regions of strong magnetic field. We find that sound speed is decreased 
in the region between approximately $r=0.99\,R_\odot$ and $r=0.995\,R_\odot$ 
(depths of 3~Mm to 7~Mm), and increased in the region between $r=0.97\,R_\odot$ 
and $r=0.985R_\odot$ (depths of 11~Mm to 21~Mm). The adiabatic index 
[$\Gamma_1$] is enhanced in the same deeper layers that sound-speed enhancement 
is seen. A weak decrease in adiabatic index is seen in the shallower layers 
in many active regions. We find that the magnitudes of these perturbations 
depend on the strength of the surface magnetic field, but we find a great 
deal of scatter in this relation, implying other factors may be relevant.
\end{abstract}
\end{opening}

\section{Introduction}
\label{sec:intro}
Measurements of the thermal structure beneath active regions using the 
techniques of local helioseismology are of substantial interest in 
solar physics. An increasingly wide variety of sophisticated models of 
the structure of sunspots and active regions are becoming available, 
and helioseismology can be used to test the validity of these models. 
In this work, we use ring diagrams to study a large number of active 
regions from Solar Cycle 23.

The use of ring diagrams \citep{Hill88} to study the near-surface 
acoustic properties, dynamics, and thermal structure of the Sun is 
by now well established \citep[e.g. review by][]{GB05}. In 
previous works, inversions of ring-diagram
frequencies for structure have been performed on small numbers of active regions 
\citep{Betal04,Bogartetal08} to determine the changes in sound speed
and adiabatic index.  In these works, sound speed and adiabatic index
were found to be enhanced in the layers between approximately $0.975\,R_\odot$
and $0.985\,R_\odot$, and depressed in the shallower layers between
$0.99\,R_\odot$ and $0.998\,R_\odot$.

In this work, we study the sub-surface structure of a much larger sample of 
active regions than have previously been examined. We proceed in two ways: 
{\it i)\/} we simply extend the work of \citet{Betal04} and \citet{Bogartetal08}, 
performing structure inversions on ring-diagram data in the same manner as 
these earlier studies, and {\it ii)} in order to better take advantage of the 
large data set that we have, we perform a Principal Component Analysis (PCA) 
and study the resulting principal components. In Section \ref{sec:data} 
we introduce the ring-diagram data sets that we use in this work and 
discuss the differences between the sample that we use for individual inversions 
and the sample that we use in our PCA. In Section \ref{sec:pca}, we describe the use of
Principal Component Analysis to study the ring-diagram mode parameters. 
In Section \ref{sec:inv}, we present the inversions of both a large  
number of individual rings and of the principal components from Section 
\ref{sec:pca}. We summarize our findings in Section \ref{sec:conc}.

\section{Data}
\label{sec:data}
\subsection{Ring Diagrams}
Ring diagrams are three-dimensional power spectra of localized regions on the Sun. In this work, the 
data are resolved line-of-sight Doppler measurements (``Dopplergrams'') taken 
with the {\it Michelson Doppler Imager\/} (MDI) instrument on the {\it Solar and Heliospheric 
Observatory\/} (SOHO). In order to achieve sufficiently high spatial frequency and 
sufficient signal-to-noise, we use only full-disk Dopplergrams, and require a 
duty cycle greater than 80\%. Because full-disk data are an MDI high-rate data 
product, sufficient data coverage is only available during yearly Dynamics 
campaigns, which are typically two or three months long. The data have a one-minute 
cadence and are tracked for 8192 minutes. The midpoint of the tracking 
interval coincides with the region of interest crossing the central meridian of 
the solar disk. The data are projected onto a rectangular grid using Postel's 
projection and corrected for the distortion in the MDI optics. This process is 
described in more detail by \citet{Patron1997} and \citet{BAT99}.

The spectra are fit in the way described by \citet{Betal04}. A model of the 
power spectrum with 13 free parameters is fit at constant frequency. This 
model accounts for advection in the zonal and meridional directions as well 
as asymmetry in the spatial frequency direction and azimuthally around the 
ring. The parameter of interest in this work is the spatial frequency [$k$], 
which is returned as a function of the temporal frequency [$\nu$] and the 
radial order [$n$]. In the plane-wave approximation the spatial frequency 
is related to the spherical harmonic degree [$\ell$] by $\ell = k R_\odot$. 
We interpolate the ring fits to integer values of $\ell$ and interpret 
them in the same way as global mode parameters. We will refer to these 
interpolated ring-diagram fits as ``mode parameter sets''; all results in 
this work are based on these data sets.

The level of activity in a ring is characterized by a Magnetic Activity Index 
\citep[MAI: ][]{Betal04}. This is a measure of the strong unsigned magnetic flux 
within the ring-diagram aperture, averaged over the tracking period.

\subsection{Ring Diagram Selection}
In this work, we use two different data sets to study the structure beneath 
active regions. The first set consists of a set of frequency differences 
between active regions from the NOAA catalog and nearby regions of quiet 
Sun, and is essentially equivalent to the data sets used in earlier works 
\citep[{\it e.g.} ][]{BAT99,Raj01,Betal04}, albeit with a much larger sample. The 
second set consists of mode frequencies for a set of largely non-overlapping 
rings, and is used for the statistical study described in Section \ref{sec:pca}.

The first sample is the same sample used by \citet{Baldner3}. 
Regions are selected from the NOAA active-region catalog between June 
1996 and April 2008. We include only NOAA active regions that are 
identified at disk center --- in other words, we require that the active 
region be present through the center of the tracking interval. We require, 
too, sufficient data coverage as noted above. As in \citet{Betal04}, we 
perform inversions relative to quiet-Sun ring-diagram mode parameters. 
The comparison ring diagrams are tracked at the same latitude as the 
active region that they are compared with. They are tracked across the disk in 
exactly the same way as the active regions, and are located within 
60$^\circ$ longitude of their active region. This is done to minimize 
the systematic errors in mode-parameter estimation arising from projection 
effects and secular changes in the instrument, as well as to minimize 
the effects of difficulties in modeling the near-surface layers of the 
solar interior. A preliminary version of this analysis was presented 
by \citet{Baldnerconf2}. Although the parent active-region sample is the 
same in both works, the actual ring-diagram data used in the inversions 
have changed somewhat --- in particular, the mode sets used and the 
comparison regions chosen. The most significant changes to the data set 
are improved treatment of the errors in the mode fitting and inversion, 
and more careful comparison region selection.

The second data set in this work is used in the PCA described in the next 
section. We use mode frequencies rather than frequency differences for 
the PCA itself. The PCA requires a filled matrix of mode parameters [$\nu_{n,\ell}$] 
for a set of modes common to all of the rings in the sample. A common set of 
[$n, \ell$] modes are chosen, and the ring-diagram fits are interpolated to 
these targets.

For the PCA procedure, we wish to minimize overlapping data, which could 
lead to some spurious statistical results. We choose only ring diagrams that 
were constructed from tracked data patches that do not overlap. The ring 
diagrams in this sample are the same size and tracking length as in the data 
set used for our individual inversions but are distributed somewhat 
differently so as to minimize spatial overlap rather than guaranteeing 
coverage of all NOAA active regions.

\section{Principal Component Analysis}
\label{sec:pca}
The use of Principal Component Analysis (PCA) in the analysis of helioseismic 
data was described by \citet{Baldner1} in the context of global-mode analysis. 
In brief, PCA finds an efficient representation of a set of data vectors as a 
linear combination of orthogonal principal components. It is efficient in the 
sense that, when properly ordered, each principal component is responsible for 
less variance in the data than the one that precedes it. In many applications, 
the data can be adequately represented using a relatively small number of 
principal components. An added benefit is that errors in measurements can be 
greatly reduced. In this work, we are primarily interested in using PCA to 
separate spurious signals in the mode parameters from those actually 
associated with solar activity, and to parametrize the changes in solar 
structure around active regions in a simple manner.

We use the following notation: if the set of all $m$ frequency measurements 
$\nu_j$ for the $i$th ring diagram form the vector $\mathbf{D}_i = \{ \nu_0 \ldots \nu_m \}$,
each observation $\mathbf{D}_i$ can be reproduced completely with the
$m$ principal components $\xi_k$ (each a vector of length $m$) and the $m$ 
scaling coefficients $c_{k,i}$ associated with that measurement:
\begin{equation}
\mathbf{D}_i = \sum_{k=0}^m c_{k,i} \xi_k.
\end{equation}
The principal components [$\xi_k$] are normalized to unity. Errors in each component
of the principal components and in the scaling coefficients are computed using a
Monte Carlo simulation.

\begin{figure}
\includegraphics[width=1\textwidth]{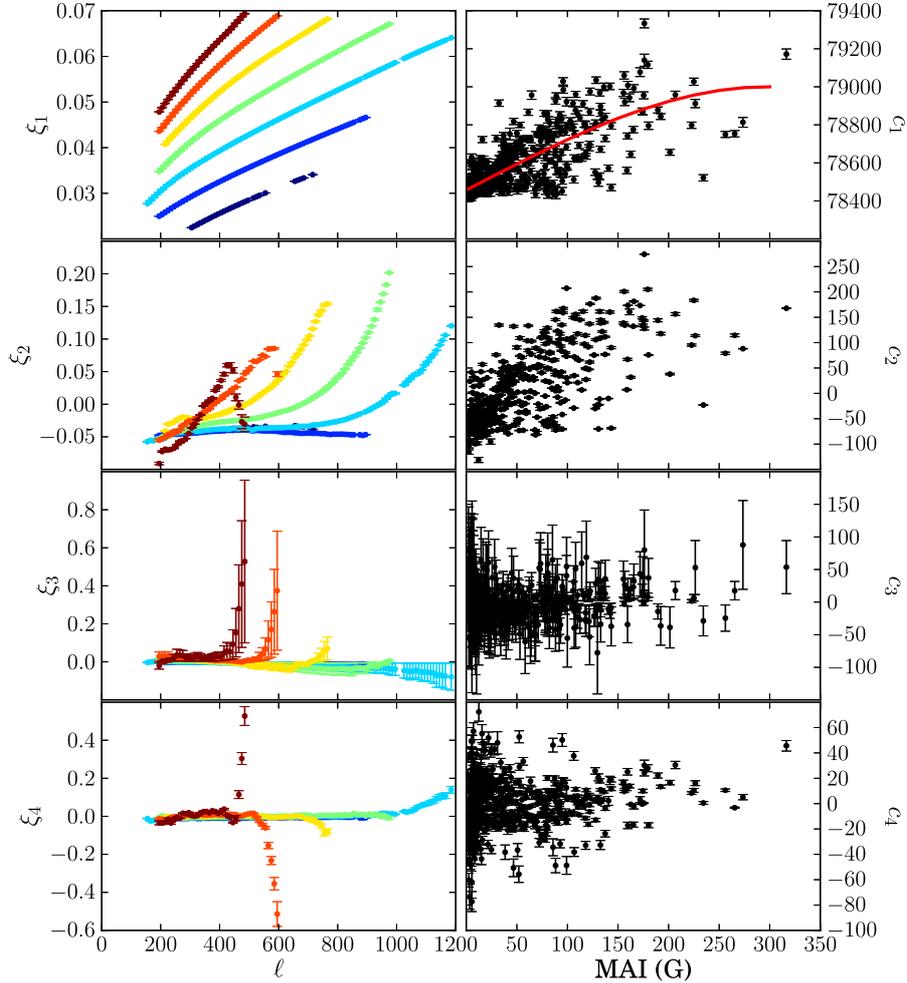}
\caption{The first four principal components and the associated scaling coefficients.
The first four principal components ($\xi_1$ through $\xi_4$) are shown in the left-hand
panels as a function of degree [$\ell$]. Each order [$n$] is shown in a different color. 
The right-hand panels show the scaling coefficients for each ring as a function of the 
ring's MAI. The red line in the top-right panel is a smoothed fit to the scaling 
coefficients.}
\label{fig:pcavec}
\end{figure}

The results of the PCA are shown in Figure \ref{fig:pcavec}. The first component may be
interpreted as the basic $\ell$\,--\,$\nu$ diagram for a ring-diagram fit. The scaling
coefficients [$c_1$] give the appropriate scaling for each observation, and we can see a
general trend with increasing magnetic activity, although the scatter is large. The second
component also has a significant dependence on magnetic activity. The following two principal
components which are shown in Figure \ref{fig:pcavec} and are dominated by tails at the
end of each $n$ ridge. These tails arise in the fits as the signal-to-noise decreases at 
the high-$\ell$ end of a ridge or the ridges begin to overlap significantly at low $\ell$. 
In either case, the fits become less reliable. In inversions on real data, we truncate the 
fits as these tails become significant.

Higher-order components are not shown --- they are in general dominated either by tails
or by small numbers of outliers. Most of these components are significant only in one
or at most a small number of rings. Some components show dependence on latitude (for
example, $\xi_5$ and $\xi_{12}$) or on epoch ($\xi_7$). By choosing to neglect
these components when reconstructing the data sets, we remove many small systematic
effects and errors from the frequency measurements.

\begin{figure}
\includegraphics[width=0.8\textwidth]{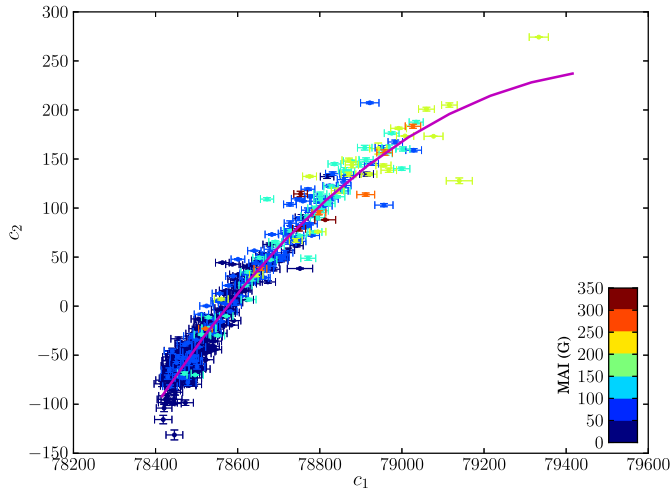}
\caption{The second principal component scaling coefficients as a function of the first.
The points are color-coded by MAI. The magenta line is a smoothed B-spline fit to the
points.}
\label{fig:coef}
\end{figure}

Given the apparent dependence of the scaling coefficients $c_1$ and $c_2$ on magnetic
activity, it is desirable to reduce these coefficients to simple functions of MAI.
In Figure \ref{fig:pcavec}, we show a smoothed fit using B-splines to the $c_1$ coefficients
as a function of MAI. These may be used to approximate the dependence of these coefficients
on magnetic activity. Since $c_1$ and $c_2$ both depend on MAI, we examine in Figure
\ref{fig:coef} their dependence on each other. It is clear that they are tightly
correlated. Thus, for a given $c_1$, which we can relate approximately to MAI, we
can determine $c_2$ fairly accurately. The scaling coefficients $c_3$ for the third
principal component do not correlate well with $c_1$ or $c_2$, or with any other
quantity that we examined.

\begin{figure}
\includegraphics[width=1\textwidth]{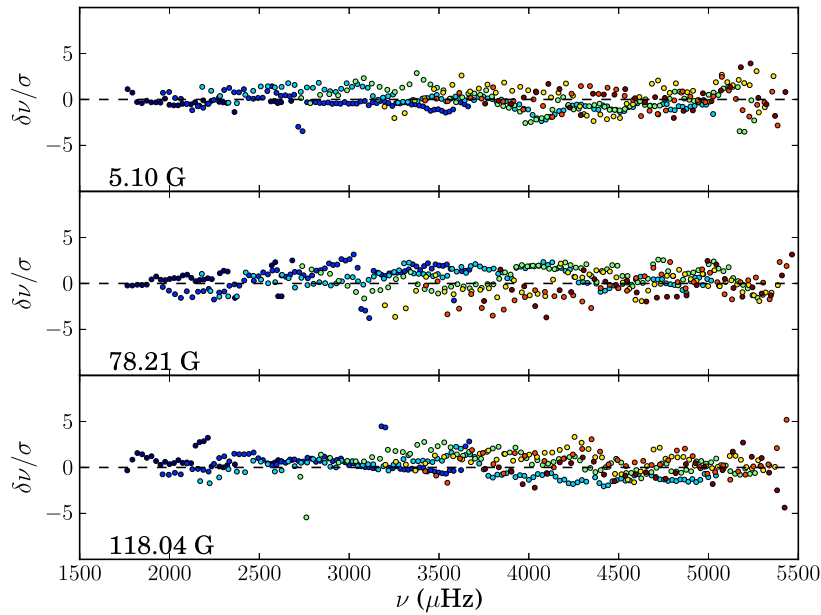}
\caption{The difference between frequencies obtained from three ring diagrams and 
their representations using the first two principal components. The residuals are 
divided by their errors. The MAIs of the three rings are shown in the lower right. 
As in Figure \ref{fig:pcavec}, different orders $n$ are plotted in different colors.}
\label{fig:resid}
\end{figure}

It is important to know how well we can reproduce the actual ring-diagram mode parameters
with a small number of vectors. In Figure \ref{fig:resid} we plot the residuals between
the reconstruction using three principal components of three randomly selected rings, and
the actual ring data. We find that, especially for low MAIs, the agreement is fairly good. At
the edges of ridges there are discrepancies, but as noted above, the spectrum fits tend to 
become unreliable at either end of the power ridges.

\section{Inversions for Structure}
\label{sec:inv}
To determine the structure of the solar layers beneath active regions (or, in general,
regions of high surface magnetic field), we invert frequency differences for sound
speed or for the first adiabatic index [$\Gamma_1$]. The treatment of this problem as
a linear inversion is by now well established both in global and in local seismology.
In this work, we use Subtractive Optimally Localized Averages \citep[SOLA:][]{PT92,PT94}
to perform the inversions.

In our SOLA inversion, there are four free parameters: the error suppression parameter
($\mu$), the cross-term trade-off parameter [$\beta$], the width of the target kernel [$\Delta_A$],
and the number of B-splines used to remove the surface term [$\Lambda$]. The selection of
these parameters has been described in earlier works, {\it e.g.\/} \citet{R-Setal99}.

\subsection{Inversions of Individual Rings}
The principal difficulty in performing an inversion for structure from helioseismic
data is in choosing the appropriate inversion parameters, which has limited 
earlier works. \citet{Baldnerconf2} found that a few combinations of 
inversion parameters worked well on a small subset of the active regions in our sample 
that was studied in detail. In this work, we again performed a thorough 
exploration of the parameter space on a small subset of the rings, including 
comparisons to rings published in earlier works. We
find that the inversion results are not strongly dependent on the choice of
surface term, so long as a surface term is used. When the surface term is not 
included in the inversions, very large perturbations are returned by the perturbations which are 
almost certainly spurious. We restrict ourselves to $\Lambda=4$. In exploring the effects of
the target kernel width, we find smaller values of $\Delta(r_0)$ tend to cause
oscillatory solutions to many inversions. Since we have not found any particularly
sharp features in our inversions, we use a fairly large value of $\Delta(r_0) = 0.055$,
which suppresses some oscillatory behavior in certain inversions, and does not
overly smooth actual structure in better inversions.

Finally, for the choices of the error suppression term and the trade-off
parameter, we have found that the values of these parameters for
acceptable inversions in the rings we have examined carefully fall in a fairly
narrow range. For the full sample, then, we run a batch of inversions for each
region over the sample. Examining the inversions for each region can then be
done fairly quickly and the best set of inversion parameters selected. We
reject regions with unstable inversions --- that is, regions whose inversion
results are very strongly dependent on the choice of inversion parameters.
What remains comprises our sample of sound speed and adiabatic index inversions.

\begin{figure}
\includegraphics[width=1\textwidth]{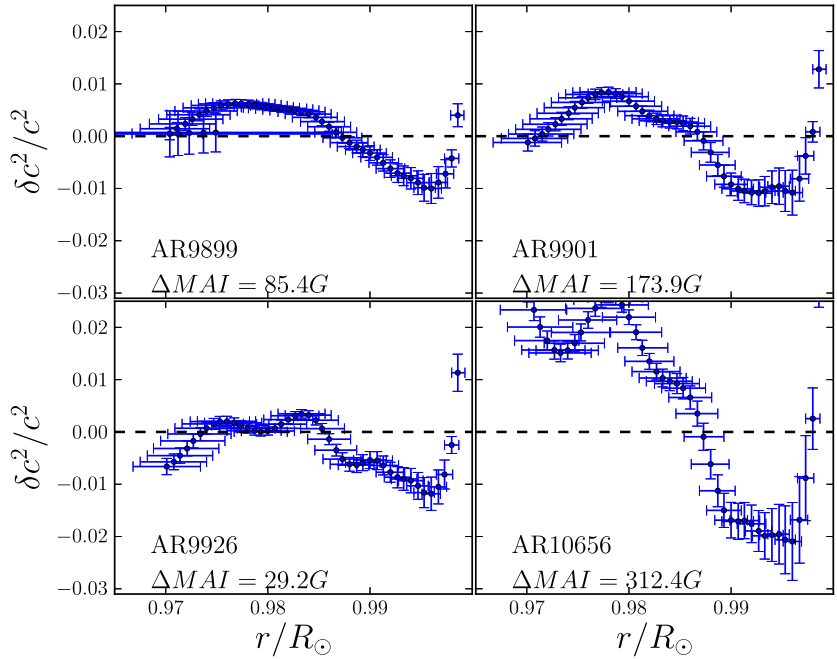}
\caption{Examples of inversions for adiabatic squared sound speed [$c^2$] for four
active regions. The sense of the inversions is active minus quiet. The plotted
horizontal error bars are the distance between the first and third quartile 
points of the inversion averaging kernels, and represent the resolution of the 
inversion. The vertical error bars are the formal errors in the inversions.}
\label{fig:c2ex}
\end{figure}

\begin{figure}
\includegraphics[width=1\textwidth]{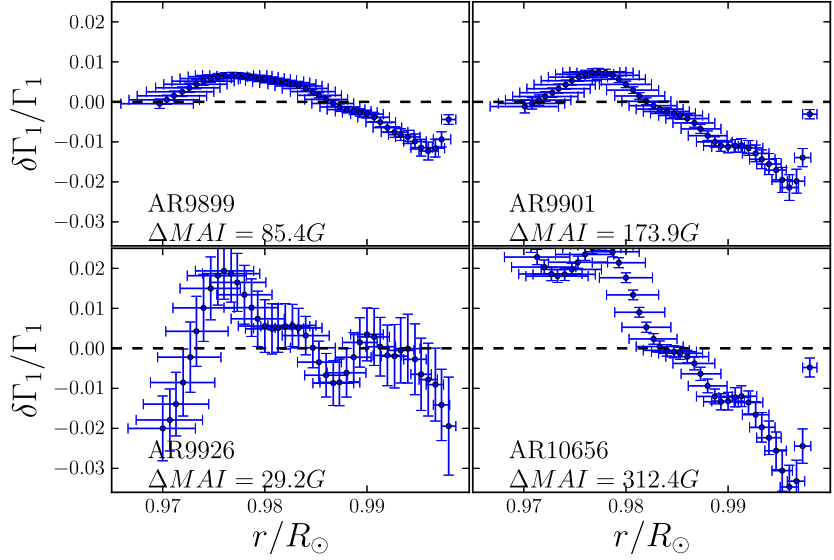}
\caption{Examples of inversions for adiabatic index [$\Gamma_1$] for four
active regions. The regions shown are the same as those in Figure \ref{fig:c2ex}.}
\label{fig:g1ex}
\end{figure}

Inversions for the difference in squared sound speed [$c^2$] and adiabatic index
[$\Gamma_1$] were performed for all regions in the sample with $\Delta$~MAI $>$ 40~G.
Figure \ref{fig:c2ex} shows example sound-speed inversions for four different
rings with a range of active-region strengths as a function of depth. Figure
\ref{fig:g1ex} shows inversions for adiabatic index for the same regions.

\begin{figure}
\includegraphics[width=1\textwidth]{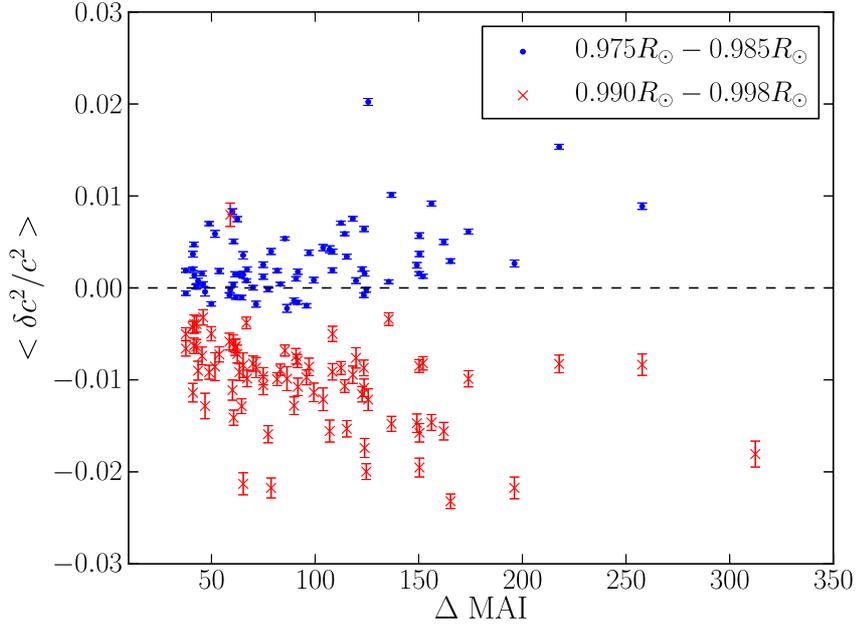}
\caption{Averages of inversions for $c^2$ over two depth ranges are shown,
plotted as a function of $\Delta$ MAI.  Blue points are averages of
inverted sound speed between $0.975\,R_\odot$ and $0.985\,R_\odot$; red crosses
are averages of inverted sound-speed difference between $0.99\,R_\odot$ and $0.998\,R_\odot$.}
\label{fig:c2}
\end{figure}

\begin{figure}
\includegraphics[width=1\textwidth]{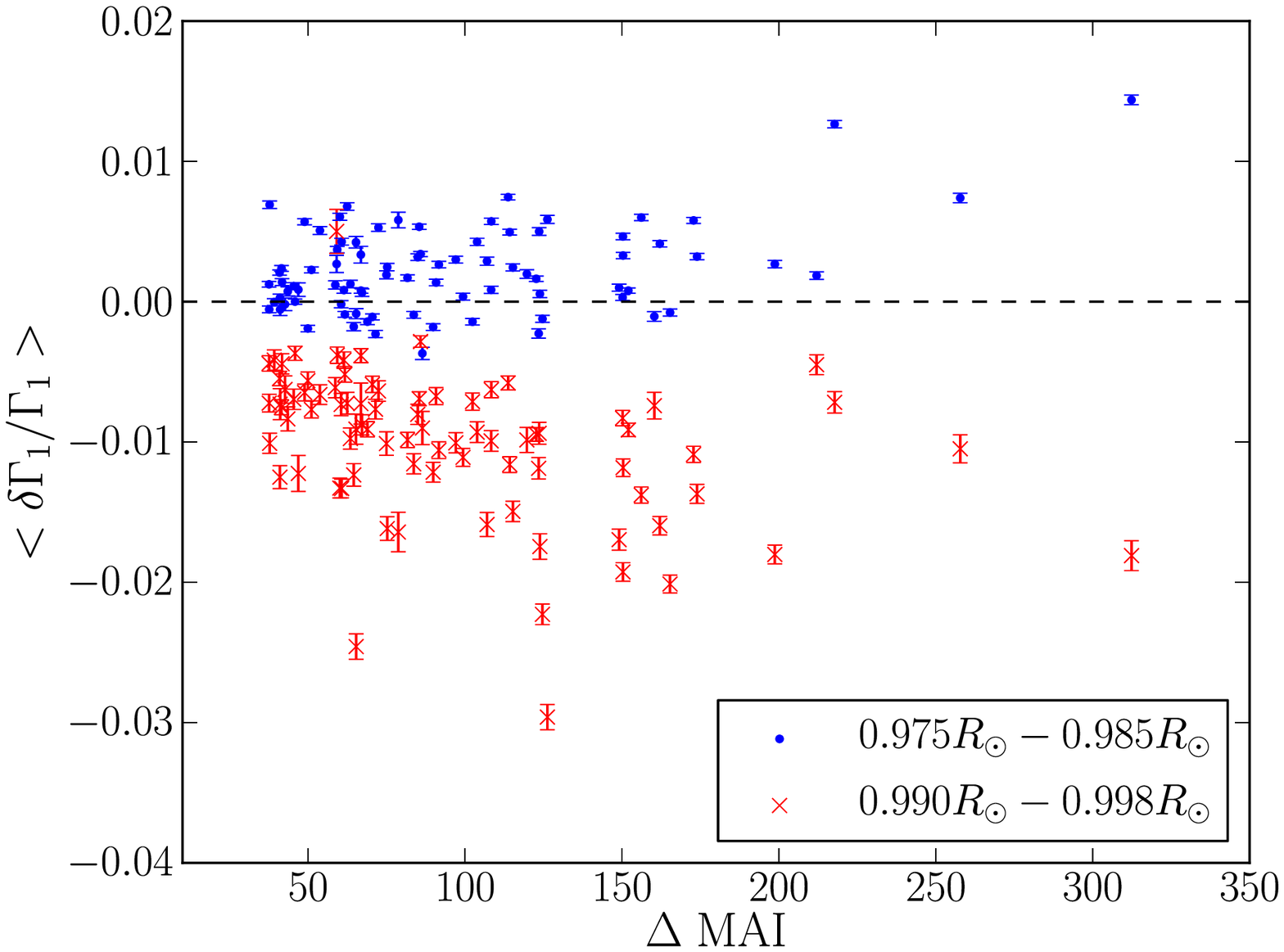}
\caption{Averages of inversions for differences in $\Gamma_1$ over two depth 
ranges are shown, plotted as a function of $\Delta$ MAI.  Blue points are averages of
inverted sound speed between $0.975\,R_\odot$ and $0.985\,R_\odot$; red crosses 
are averages of inverted sound speed between $0.99\,R_\odot$ and $0.998\,R_\odot$.}
\label{fig:g1}
\end{figure}

Figure \ref{fig:c2} shows averages of the inverted sound speed for all
regions in our sample for different depth ranges.  For sound speed
averaged between $0.975R_\odot$ and $0.985R_\odot$, sound speeds are
generally enhanced in the presence of magnetic fields, while in the
region from $0.99R_\odot$ to $0.998R_\odot$, sound speeds decrease.
In both regions, the magnitude of the change tends to increase with
magnetic-field strength, although the relationship seems to be more of
an envelope than a linear relation.  Further, there seems to be some
saturation of the effect at very high magnetic-field strengths.

In the shallowest layers ($r > 0.998\,R_\odot$), the sound-speed inversion 
results show a sharp positive change. This feature is present in every 
inversion in our sample. Caution should be used in interpreting this, 
however, as the averaging kernels become strongly asymmetric near the 
surface and the cross-term contributions begin to become significant. 
Examples of this sharp feature can be seen in the inversions in 
Figure \ref{fig:c2ex}.

We have also inverted for the first adiabatic index. In
Figure \ref{fig:g1}, we show the same averages as Figure \ref{fig:c2},
but for the $\Gamma_1$ inversions. In general, $\Gamma_1$ is an
easier quantity to invert for (in the sense that the inversions
tend to be less sensitive to the inversion parameters), and so
we include a larger number of inversions for $\Gamma_1$ than we
could for $c^2$.

We find that, as for the $c^2$ results, there is a depression in $\Gamma_1$
in the shallower layers that we invert (above $r=0.99\,R_\odot$), and
that, in many cases, there is a corresponding enhancement below
approximately $r=0.98\,R_\odot$, as was found in earlier works. We
find that the deeper enhancement is, for most rings, much less pronounced
than for the $c^2$. For some regions, in fact, we do not see any
positive perturbation at all, and in general we find only a weak
correlation with magnetic activity.

\begin{figure}
\centerline{\hspace*{0.015\textwidth}
\includegraphics[width=0.515\textwidth]{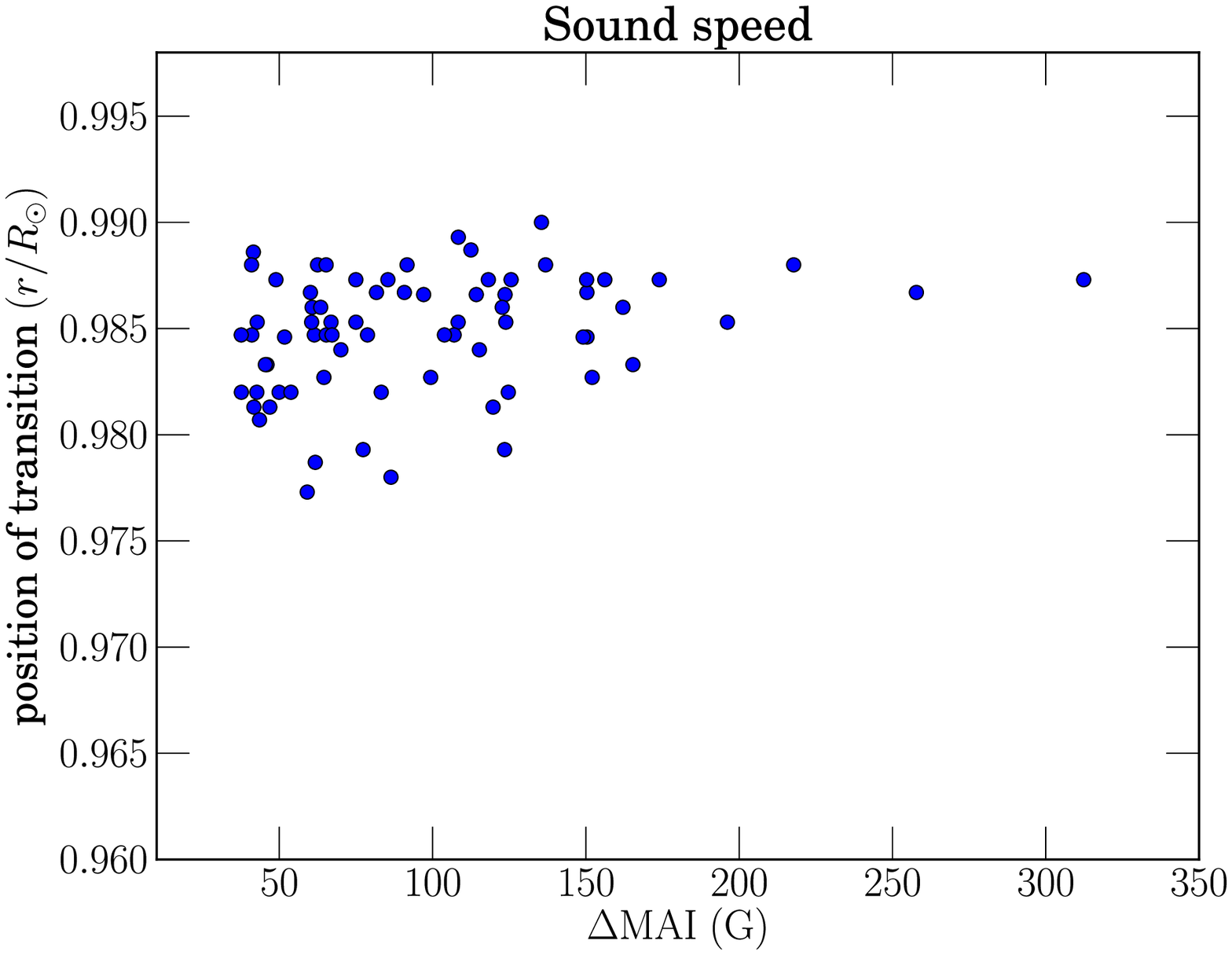}
\hspace*{-0.03\textwidth}
\includegraphics[width=0.515\textwidth]{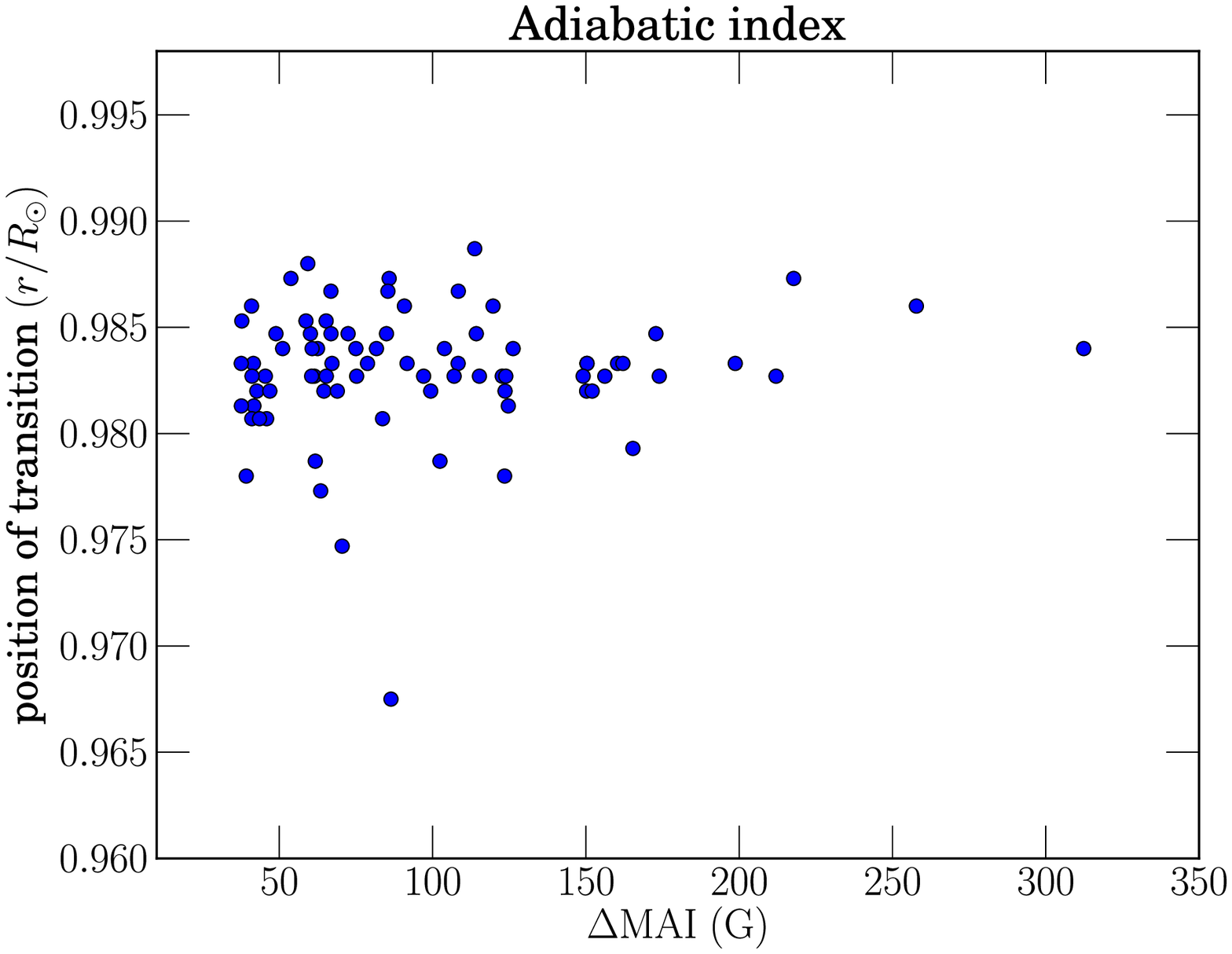}
}
\caption{Position of transition, in fractional radius, between the negative
(shallower) inversion results and the deeper (positive) perturbations in the 
inversion results, plotted as a function of $\Delta$ MAI. At left, the sound 
speed results are shown, at right are the results for adiabatic index. These 
transition points are found by treating the inversion points as a continuous 
curve and finding the zero-crossing point.}
\label{fig:crossc2}
\end{figure}

The depth ranges of the negative and positive perturbations are found 
to be relatively constant. In Figure \ref{fig:crossc2}, we plot the 
depths at which our inversion results change from negative to 
positive. We find that, at lower activity levels, there is greater 
variance in the depths of the perturbations, but this can be 
explained by lower signal-to-noise in the inversions. Beyond that, 
we do not find any significant change in the depth ranges of the 
perturbations.

\subsection{Inversions of PCA Data}
\label{sec:invpca}
In this section we use the PCA reconstruction of ring-diagram mode parameters
described in Section \ref{sec:pca} to perform inversions for structure. As
discussed earlier, the advantages of using PCA on this data set are reduced
errors in the reconstructed data set and removal of certain systematic
effects in parameter estimation due to secular changes in the MDI instrument
over time and projection effects. In addition, because we may reduce the
data to a number of linearly independent vectors, we substantially reduce 
the number of inversions that need to be done.

\begin{figure}
\includegraphics[width=1\textwidth]{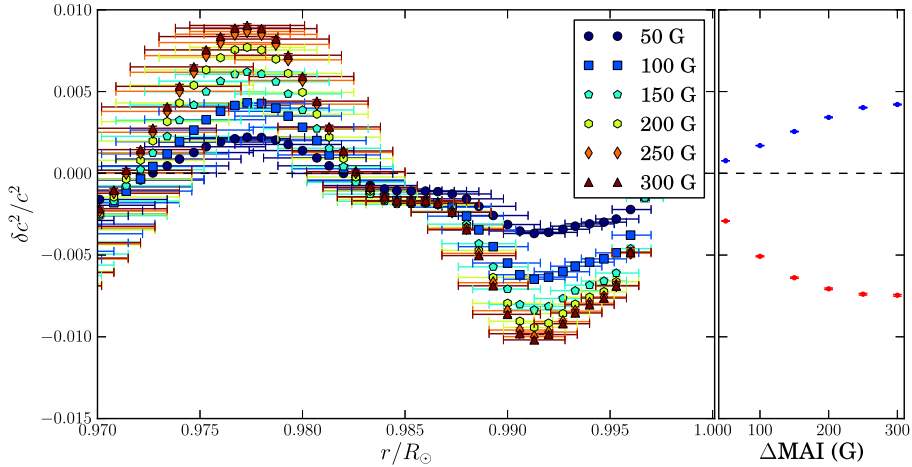}
\caption{Inversion of the first two principal components at different MAIs for $c^2$. 
The scaling coefficients are chosen as discussed in the text. The left-hand panel 
shows individual inversions as a function of radius. As in 
Figure \ref{fig:c2ex}, the horizontal error bars are the difference between the first 
and third quartile points of the averaging kernels, and represent the resolution of 
the inversion. The vertical error bars are the formal errors in the inverted quantities. 
The latter are generally smaller than the points. The right-hand panel shows the average 
of the inversion results over the same depth ranges as Figure \ref{fig:c2} as a function 
of MAI.}
\label{fig:pcacsq}
\end{figure}

Most of the variation in the ring-diagram mode parameter sets are spanned 
by the first two principal components (see Figure \ref{fig:resid}). 
Furthermore, as the coefficients $c_1$ are relatively well correlated with 
MAI, and $c_2$ is tightly correlated with $c_1$, we can reconstruct ring-diagram 
frequencies as a function of MAI. We then compute frequencies [$\nu_{n,\ell}$],
the difference in frequency [$\delta \nu_{n,\ell}$] between the target MAI and
zero MAI, and the error in $\delta \nu_{n,\ell}$. With these quantities, we
may perform a SOLA inversion in the same manner as in the previous section.

The inversion results for six target MAIs between 50~G and 300~G are shown in
Figure \ref{fig:pcacsq}. As in the inversions of individual rings, the inversions
of the PCA reconstructions show a distinctive two-layer structure with a shallow
negative perturbation and a deeper positive perturbation. In addition, the
magnitudes of these perturbations scale with MAI, and the saturation effect
hinted at in the results in Figure \ref{fig:c2} is clearly present here.

While most of the variation is spanned by $\xi_1$ and $\xi_2$, there remains 
significant signal in subsequent components. We are interested to know, then, 
what effects these higher components have on inversions for structure, and so 
we invert components separately. It must be noted, however, that inversion 
results from different components, each with different errors, cannot be 
simply combined together linearly. With that in mind, we proceed to invert 
various principal components separately.

\begin{figure}
\includegraphics[width=1\textwidth]{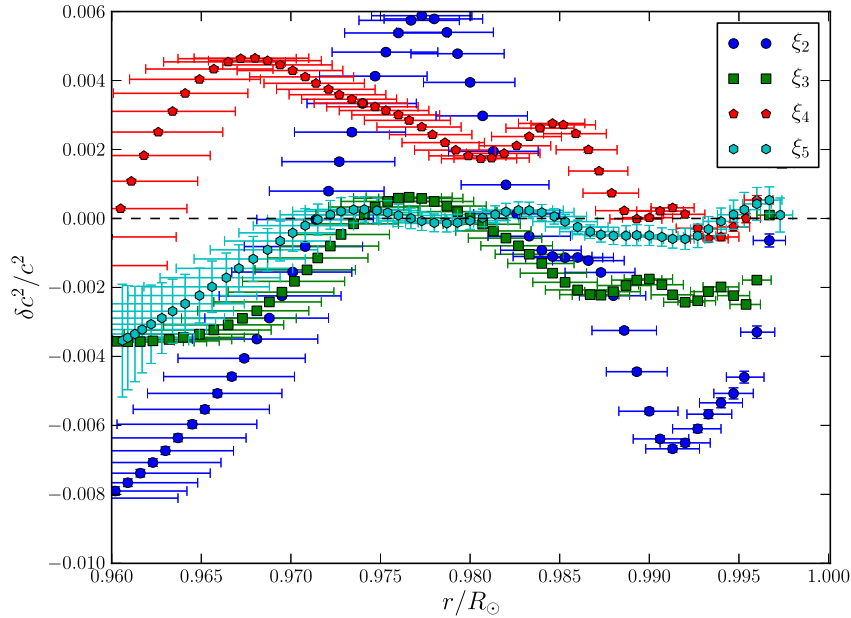}
\caption{Inversion of the $\xi_2$ through $\xi_5$ principal components individually. 
Each component $\xi_i$ is scaled by the largest associated coefficient $c_i$ in our 
data set. The base frequencies are the same as the base frequencies used in 
Figure \ref{fig:pcacsq}.}
\label{fig:vecinv}
\end{figure}

In Figure \ref{fig:vecinv}, we show inversions for sound speed for 
principal components $\xi_2$ through $\xi_5$. Because it is not clear 
how to parametrize the scaling coefficients for any of the components 
except $\xi_2$, we simply scale the component by the largest 
coefficient for that component, and perform the inversion. We invert 
the components themselves, not explicit differences --- thus the 
inverted sound speeds are the difference between a ring with a large 
coefficient $c_i$ and a ring without any contribution from that 
component at all. We see that $\xi_3$ and $\xi_4$ do
return changes in structure, but $\xi_5$ is consistent with no change.

\begin{figure}
\includegraphics[width=1\textwidth]{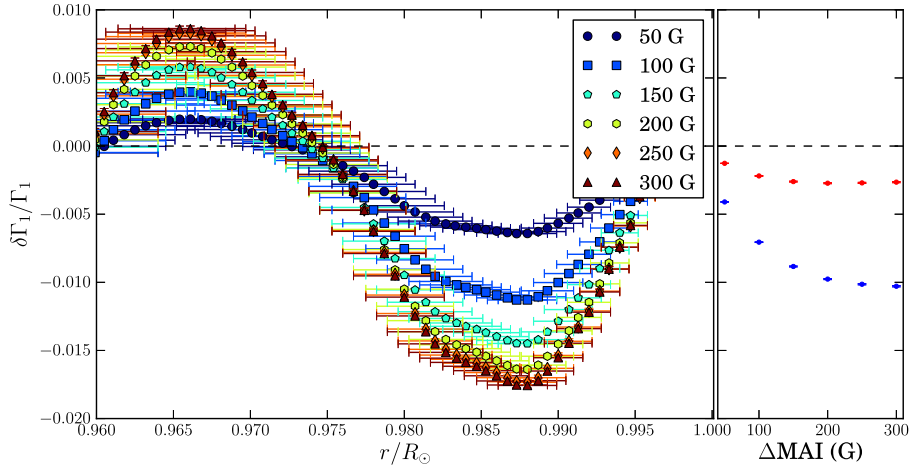}
\caption{Inversion of the first two principal components at different MAIs for
adiabatic index [$\Gamma_1$]. The data are the same as those inverted in Figure 
\ref{fig:pcacsq}. The left hand panel shows the individual inversions, the 
right hand panel shows averages over the same depth ranges as Figure 
\ref{fig:g1} as a function of MAI.}
\label{fig:pcag1}
\end{figure}

We also invert for adiabatic index. Figure \ref{fig:pcag1} shows
inversions for the first two principal components at various MAIs between
50~G and 300~G. We find general consistency with the individual ring-diagram
inversions: a shallow negative and deeper positive perturbation. The 
correlation with MAI in the shallow layer is very weak, and in some rings 
there is no negative perturbation at all. We find that the boundary between 
the negative and positive perturbations is somewhat deeper than in 
the individual inversion for $\Gamma_1$, occurring between $0.975\,R_\odot$ 
and $0.98\,R_\odot$. This discrepancy does not appear to arise from 
the choice of modes used in the inversion, nor does it depend on the choice 
of inversion parameters. One possibility is that the contributions from 
other principal components may shift the boundary upward, which would in 
turn imply a dependence of $\Gamma_1$ on something other than MAI. 
This is in fact what we find. On examination, the $\xi_4$ principal 
component returns a positive perturbation in the region we expect, and 
little signal elsewhere. The magnitude of the $\xi_4$ component (the $c_4$ 
coefficients from Figure \ref{fig:pcavec}) show a large scatter at low MAI 
values, but a pronounced bump above approximately 150G. This is broadly 
consistent with the individual inversion results shown in Figure \ref{fig:g1}, 
where many of the regions with $\Delta$ MAI $<$ 150G do not show positive 
perturbations.

We have 
explored the first ten principal components in some detail and not found 
evidence of this, however. 

\section{Conclusions}
\label{sec:conc}
In this work, we have presented inversions for structure of a large number of
active regions, and used Principal Component Analysis to parametrize more 
accurately these inversion results. We confirm earlier results using 
ring-diagram analysis \citep{Betal04,Bogartetal08}, as well as high-degree 
global-mode analysis \citep{CRS2012} that finds a two-layer thermal structure beneath
surface magnetic activity. We find a shallow negative sound speed perturbation
and a somewhat deeper positive perturbation, with a similar structure in the
adiabatic index. The magnitudes of these perturbations generally
increase with the magnitude of the surface magnetic field.

\citet{Bogartetal08} found a linear correlation between magnitude of
the sound-speed change and strength of the active region.  We find
a similar relation, but with substantial scatter in our inversions of
individual rings. It is perhaps more correct to say that we find 
an ``envelope'' related to MAI than a real correlation. Further, we find
that the correlation appears to saturate at high field strengths.
We find that the positive perturbation in $\Gamma_1$ is much weaker 
than in $c^2$, and that the correlation with magnetic activity is much 
less significant. We find consistent values for the strength of 
the perturbations in the shallower layers with \citet{Betal04} and 
with \citet{Bogartetal08}, 
but we find consistently smaller values for the deeper, positive 
perturbations than those found in the previous works. The choices of 
inversion parameters and mode sets can have some effect on the 
magnitudes of the inversions. In \citet{Betal04}, RLS inversions tended 
to return slightly larger perturbations than the SOLA inversions, which we 
employ in this work. The most significant effect on our results appears 
to be our decision to neglect the $f$-mode in our inversions, which, as 
noted above, made our inversion results more stable. It also appears to 
have decreased the magnitude of the sound-speed change in the deeper 
layers.

\citet{Bogartetal08} also reported that the depths of the positive
perturbations in $\Gamma_1$ were deeper than those in the $c^2$
inversions. We find this unambiguously in the PCA inversion results
(Figure \ref{fig:pcag1}), and it is also consistent with the inversions
of individual rings.

Interpreting these results is not straightforward. The inversions that we perform in 
this work are, strictly speaking, only valid in a spherically symmetric, 
non-magnetized star. Clearly, neither of these conditions is satisfied 
in active regions. In the presence of magnetic fields, wave propagation 
is affected both by the direct effect of the Lorentz force, and by the 
effects of magnetic fields on the thermal structure. At some point --- at 
some minimum field strength or field configuration ---  the entire 
linearized inversion becomes meaningless, as the contributions from 
magnetic fields to the wave propagation (which is, among other problems, 
no longer insensitive to direction) become significant.

In some studies, the 
choice is made to interpret the inferred change in sound speed as 
a change in the local wave speed, 
as was done in \citet{Gizonetal09} and \citet{Moradietal10}, rather than interpreting it 
as the effects of a thermal perturbation as we do here. In these works, 
significant disagreement was found when ring-diagram inversions were 
compared to time--distance inversions for the same solar data. The resolution 
of this discrepancy remains unclear. If we do in fact measure a 
wave speed perturbation in this work, \citet{Linetal09} have claimed that 
the magnetic effect can be disentangled from the thermal effect. In the 
case of time--distance analysis, \citet{Braunetal12} have shown that travel-time 
shifts determined from a model sunspot match neither what the thermal 
perturbation should give nor what they would expect from the magnetoacoustic 
fast mode speed. The implication of this work is that, in the case of time--distance 
analysis, at least, it is not correct to interpret inversion results as either 
changes in the thermal sound speed or as local wave speed perturbations. Equivalent 
work has not yet been done for ring-diagram analysis, but must be done to determine 
the extent to which the assumptions we have made are valid.

In studying our sample of rings using PCA, we have decomposed the ring
diagram frequency measurements into a set of linearly independent components.
The dependence of the first two of these on magnetic activity allows us to
parametrize most of the frequency variance across our sample of rings as a
function of MAI, and to determine what changes in sound speed and
adiabatic index could give rise to these changes. We find that both
sound speed and adiabatic index have a dependence on MAI that is consistent
with what has been found in individual rings.

There is further variance in the ring-diagram frequency measurements, however,
and MAI alone does not adequately parametrize the ring-diagram frequencies, as
can be seen in Figure \ref{fig:pcavec}. Many of these changes are due to
errors in the ring fitting and systematic effects due to projection effects
and secular changes in the MDI instrument, but some of the more significant
principal components may represent changes that are solar in origin and that
may be associated with thermal changes below the solar surface, as shown in
Figure \ref{fig:vecinv}.

We have demonstrated that PCA can be a useful tool in ring-diagram analysis
and structure inversion. The large volume of data being returned from the
{\it Helioseismic and Magnetic Imager\/} (HMI) on the {\it Solar Dynamics Observatory\/} (SDO)
spacecraft represents a significant data analysis challenge --- it is
possible that this technique might prove a feasible alternative to attempting
inversions on every ring diagram produced by HMI.

\begin{acks}
This work was partially supported by a NASA Earth and Space Sciences fellowship 
NNX08AY41H to CSB. CSB and RSB are currently supported by NASA grant NAS5-02139 
to Stanford University. SB acknowledges support from NASA grant NNX10AE60G. This 
work utilizes data from the {\it Solar Oscillations 
Investigation/Michelson Doppler Imager\/} (SOI/MDI) on the {\it Solar and Heliospheric 
Observatory\/} (SOHO).  SOHO is a project of international cooperation
between ESA and NASA.  MDI is supported by NASA grant NNX09AI90G to Stanford 
University.
\end{acks}

%%\bibliographystyle{spr-mp-sola}
%%\bibliography{bibfile}

\begin{thebibliography}{18}
% BibTex style file: spr-mp-sola.bst (nameyear), 2011-09-16
\ifx \bisbn   \undefined \def \bisbn  #1{ISBN #1}\fi
\ifx \binits  \undefined \def \binits#1{#1}\fi
\ifx \bauthor  \undefined \def \bauthor#1{#1}\fi
\ifx \batitle  \undefined \def \batitle#1{#1}\fi
\ifx \bjtitle  \undefined \def \bjtitle#1{\textit{#1}}\fi
\ifx \bvolume  \undefined \def \bvolume#1{\textbf{#1}}\fi
\ifx \byear  \undefined \def \byear#1{#1}\fi
\ifx \bissue  \undefined \def \bissue#1{#1}\fi
\ifx \bfpage  \undefined \def \bfpage#1{#1}\fi
\ifx \blpage  \undefined \def \blpage #1{#1}\fi
\ifx \burl  \undefined \def \burl#1{\textsf{#1}}\fi
\ifx \href  \undefined \def \href#1#2{\textsf{#2}}\fi
\ifx \doiurl  \undefined \def
  \doiurl#1{\href{http://dx.doi.org/#1}{\textsf{#1}}}\fi
\ifx \betal  \undefined \def \betal{\textit{et al.}}\fi
\ifx \binstitute  \undefined \def \binstitute#1{#1}\fi
\ifx \bctitle  \undefined \def \bctitle#1{#1}\fi
\ifx \beditor  \undefined \def \beditor#1{#1}\fi
\ifx \bpublisher  \undefined \def \bpublisher#1{#1}\fi
\ifx \bbtitle  \undefined \def \bbtitle#1{\textit{#1}}\fi
\ifx \bedition  \undefined \def \bedition#1{#1}\fi
\ifx \bseriesno  \undefined \def \bseriesno#1{\textbf{#1}}\fi
\ifx \blocation  \undefined \def \blocation#1{#1}\fi
\ifx \bsertitle  \undefined \def \bsertitle#1{\textit{#1}}\fi
\ifx \bsnm \undefined \def \bsnm#1{#1}\fi
\ifx \bsuffix \undefined \def \bsuffix#1{#1}\fi
\ifx \bparticle \undefined \def \bparticle#1{#1}\fi
\ifx \barticle \undefined \def \barticle#1{}\fi
\ifx \botherref \undefined \def \botherref#1{}\fi
\ifx \url \undefined \def \url#1{\textsf{#1}}\fi
\ifx \bchapter \undefined \def \bchapter#1{}\fi
\ifx \bbook \undefined \def \bbook#1{}\fi
\ifx \bcomment \undefined \def \bcomment#1{#1}\fi
\ifx \oauthor \undefined \def \oauthor#1{#1}\fi
\ifx \citeauthoryear \undefined \def \citeauthoryear#1{#1}\fi
\def \endbibitem {}
\ifx \bconflocation  \undefined \def \bconflocation#1{#1} \fi

\bibitem[\protect\citeauthoryear{{Baldner} and {Basu}}{2008}]{Baldner1}
\begin{barticle}
\bauthor{\bsnm{{Baldner}}, \binits{C.S.}},
\bauthor{\bsnm{{Basu}}, \binits{S.}}:
\byear{2008},
\batitle{{Solar Cycle Related Changes at the Base of the Convection Zone}}.
\bjtitle{\apj}
\bvolume{686},
\bfpage{1349}\,--\,\blpage{1361}.
doi:\doiurl{10.1086/591514}.
\end{barticle}
\endbibitem

\bibitem[\protect\citeauthoryear{{Baldner}, {Bogart}, and
  {Basu}}{2011a}]{Baldner3}
\begin{barticle}
\bauthor{\bsnm{{Baldner}}, \binits{C.S.}},
\bauthor{\bsnm{{Bogart}}, \binits{R.S.}},
\bauthor{\bsnm{{Basu}}, \binits{S.}}:
\byear{2011}a,
\batitle{{Evidence for Solar Frequency Dependence on Sunspot Type}}.
\bjtitle{\apjl}
\bvolume{733},
\bfpage{5}.
doi:\doiurl{10.1088/2041-8205/733/1/L5}.
\end{barticle}
\endbibitem

\bibitem[\protect\citeauthoryear{{Baldner}, {Bogart}, and
  {Basu}}{2011b}]{Baldnerconf2}
\begin{barticle}
\bauthor{\bsnm{{Baldner}}, \binits{C.S.}},
\bauthor{\bsnm{{Bogart}}, \binits{R.S.}},
\bauthor{\bsnm{{Basu}}, \binits{S.}}:
\byear{2011}b,
\batitle{{The thermal structure of sunspots from ring diagram analysis}}.
\bjtitle{{\it J. Phys. Conf. Ser.\/}}
\bvolume{271}(\bissue{1}),
\bfpage{012006}.
doi:\doiurl{10.1088/1742-6596/271/1/012006}.
\end{barticle}
\endbibitem

\bibitem[\protect\citeauthoryear{{Basu}, {Antia}, and {Bogart}}{2004}]{Betal04}
\begin{barticle}
\bauthor{\bsnm{{Basu}}, \binits{S.}},
\bauthor{\bsnm{{Antia}}, \binits{H.M.}},
\bauthor{\bsnm{{Bogart}}, \binits{R.S.}}:
\byear{2004},
\batitle{{Ring-Diagram Analysis of the Structure of Solar Active Regions}}.
\bjtitle{\apj}
\bvolume{610},
\bfpage{1157}\,--\,\blpage{1168}.
doi:\doiurl{10.1086/421843}.
\end{barticle}
\endbibitem

\bibitem[\protect\citeauthoryear{{Basu}, {Antia}, and {Tripathy}}{1999}]{BAT99}
\begin{barticle}
\bauthor{\bsnm{{Basu}}, \binits{S.}},
\bauthor{\bsnm{{Antia}}, \binits{H.M.}},
\bauthor{\bsnm{{Tripathy}}, \binits{S.C.}}:
\byear{1999},
\batitle{{Ring Diagram Analysis of Near-Surface Flows in the Sun}}.
\bjtitle{\apj}
\bvolume{512},
\bfpage{458}\,--\,\blpage{470}.
doi:\doiurl{10.1086/306765}.
\end{barticle}
\endbibitem

\bibitem[\protect\citeauthoryear{{Bogart} \textit{et~al.}}{2008}]{Bogartetal08}
\begin{barticle}
\bauthor{\bsnm{{Bogart}}, \binits{R.S.}},
\bauthor{\bsnm{{Basu}}, \binits{S.}},
\bauthor{\bsnm{{Rabello-Soares}}, \binits{M.C.}},
\bauthor{\bsnm{{Antia}}, \binits{H.M.}}:
\byear{2008},
\batitle{{Probing the Subsurface Structures of Active Regions with Ring-Diagram
  Analysis}}.
\bjtitle{\solphys}
\bvolume{251},
\bfpage{439}\,--\,\blpage{451}.
doi:\doiurl{10.1007/s11207-008-9213-9}.
\end{barticle}
\endbibitem

\bibitem[\protect\citeauthoryear{{Braun} \textit{et~al.}}{2012}]{Braunetal12}
\begin{barticle}
\bauthor{\bsnm{{Braun}}, \binits{D.C.}},
\bauthor{\bsnm{{Birch}}, \binits{A.C.}},
\bauthor{\bsnm{{Rempel}}, \binits{M.}},
\bauthor{\bsnm{{Duvall}}, \binits{T.L.}}:
\byear{2012},
\batitle{{Helioseismology of a Realistic Magnetoconvective Sunspot
  Simulation}}.
\bjtitle{\apj}
\bvolume{744},
\bfpage{77}.
doi:\doiurl{10.1088/0004-637X/744/1/77}.
\end{barticle}
\endbibitem

\bibitem[\protect\citeauthoryear{{Gizon} and {Birch}}{2005}]{GB05}
\begin{barticle}
\bauthor{\bsnm{{Gizon}}, \binits{L.}},
\bauthor{\bsnm{{Birch}}, \binits{A.C.}}:
\byear{2005},
\batitle{{Local Helioseismology}}.
\bjtitle{Living Reviews in Solar Physics}
\bvolume{2},
No.1 (\url{http://www.livingreviews.org/lrsp-2005-6})
\end{barticle}
\endbibitem

\bibitem[\protect\citeauthoryear{{Gizon} \textit{et~al.}}{2009}]{Gizonetal09}
\begin{barticle}
\bauthor{\bsnm{{Gizon}}, \binits{L.}},
\bauthor{\bsnm{{Schunker}}, \binits{H.}},
\bauthor{\bsnm{{Baldner}}, \binits{C.S.}},
\bauthor{\bsnm{{Basu}}, \binits{S.}},
\bauthor{\bsnm{{Birch}}, \binits{A.C.}},
\bauthor{\bsnm{{Bogart}}, \binits{R.S.}},
\bauthor{\bsnm{{Braun}}, \binits{D.C.}},
\bauthor{\bsnm{{Cameron}}, \binits{R.}},
\bauthor{\bsnm{{Duvall}}, \binits{T.L.}},
\bauthor{\bsnm{{Hanasoge}}, \binits{S.M.}},
\bauthor{\bsnm{{Jackiewicz}}, \binits{J.}},
\bauthor{\bsnm{{Roth}}, \binits{M.}},
\bauthor{\bsnm{{Stahn}}, \binits{T.}},
\bauthor{\bsnm{{Thompson}}, \binits{M.J.}},
\bauthor{\bsnm{{Zharkov}}, \binits{S.}}:
\byear{2009},
\batitle{{Helioseismology of Sunspots: A Case Study of NOAA Region 9787}}.
\bjtitle{\ssr}
\bvolume{144},
\bfpage{249}\,--\,\blpage{273}.
doi:\doiurl{10.1007/s11214-008-9466-5}.
\end{barticle}
\endbibitem

\bibitem[\protect\citeauthoryear{{Hill}}{1988}]{Hill88}
\begin{barticle}
\bauthor{\bsnm{{Hill}}, \binits{F.}}:
\byear{1988},
\batitle{{Rings and trumpets - Three-dimensional power spectra of solar
  oscillations}}.
\bjtitle{\apj}
\bvolume{333},
\bfpage{996}\,--\,\blpage{1013}.
doi:\doiurl{10.1086/166807}.
\end{barticle}
\endbibitem

\bibitem[\protect\citeauthoryear{{Lin}, {Basu}, and {Li}}{2009}]{Linetal09}
\begin{barticle}
\bauthor{\bsnm{{Lin}}, \binits{C.-H.}},
\bauthor{\bsnm{{Basu}}, \binits{S.}},
\bauthor{\bsnm{{Li}}, \binits{L.}}:
\byear{2009},
\batitle{{Interpreting Helioseismic Structure Inversion Results of Solar Active
  Regions}}.
\bjtitle{\solphys}
\bvolume{257},
\bfpage{37}\,--\,\blpage{60}.
doi:\doiurl{10.1007/s11207-009-9332-y}.
\end{barticle}
\endbibitem

\bibitem[\protect\citeauthoryear{{Moradi} \textit{et~al.}}{2010}]{Moradietal10}
\begin{barticle}
\bauthor{\bsnm{{Moradi}}, \binits{H.}},
\bauthor{\bsnm{{Baldner}}, \binits{C.}},
\bauthor{\bsnm{{Birch}}, \binits{A.C.}},
\bauthor{\bsnm{{Braun}}, \binits{D.C.}},
\bauthor{\bsnm{{Cameron}}, \binits{R.H.}},
\bauthor{\bsnm{{Duvall}}, \binits{T.L.}},
\bauthor{\bsnm{{Gizon}}, \binits{L.}},
\bauthor{\bsnm{{Haber}}, \binits{D.}},
\bauthor{\bsnm{{Hanasoge}}, \binits{S.M.}},
\bauthor{\bsnm{{Hindman}}, \binits{B.W.}},
\bauthor{\bsnm{{Jackiewicz}}, \binits{J.}},
\bauthor{\bsnm{{Khomenko}}, \binits{E.}},
\bauthor{\bsnm{{Komm}}, \binits{R.}},
\bauthor{\bsnm{{Rajaguru}}, \binits{P.}},
\bauthor{\bsnm{{Rempel}}, \binits{M.}},
\bauthor{\bsnm{{Roth}}, \binits{M.}},
\bauthor{\bsnm{{Schlichenmaier}}, \binits{R.}},
\bauthor{\bsnm{{Schunker}}, \binits{H.}},
\bauthor{\bsnm{{Spruit}}, \binits{H.C.}},
\bauthor{\bsnm{{Strassmeier}}, \binits{K.G.}},
\bauthor{\bsnm{{Thompson}}, \binits{M.J.}},
\bauthor{\bsnm{{Zharkov}}, \binits{S.}}:
\byear{2010},
\batitle{{Modeling the Subsurface Structure of Sunspots}}.
\bjtitle{\solphys}
\bvolume{267},
\bfpage{1}\,--\,\blpage{62}.
doi:\doiurl{10.1007/s11207-010-9630-4}.
\end{barticle}
\endbibitem

\bibitem[\protect\citeauthoryear{{Patron} \textit{et~al.}}{1997}]{Patron1997}
\begin{barticle}
\bauthor{\bsnm{{Patron}}, \binits{J.}},
\bauthor{\bsnm{{Gonzalez Hernandez}}, \binits{I.}},
\bauthor{\bsnm{{Chou}}, \binits{D.-Y.}},
\bauthor{\bsnm{{Sun}}, \binits{M.-T.}},
\bauthor{\bsnm{{Mu}}, \binits{T.-M.}},
\bauthor{\bsnm{{Loudagh}}, \binits{S.}},
\bauthor{\bsnm{{Bala}}, \binits{B.}},
\bauthor{\bsnm{{Chou}}, \binits{Y.-P.}},
\bauthor{\bsnm{{Lin}}, \binits{C.-H.}},
\bauthor{\bsnm{{Huang}}, \binits{I.-J.}},
\bauthor{\bsnm{{Jimenez}}, \binits{A.}},
\bauthor{\bsnm{{Rabello-Soares}}, \binits{M.C.}},
\bauthor{\bsnm{{Ai}}, \binits{G.}},
\bauthor{\bsnm{{Wang}}, \binits{G.-P.}},
\bauthor{\bsnm{{Zirin}}, \binits{H.}},
\bauthor{\bsnm{{Marquette}}, \binits{W.}},
\bauthor{\bsnm{{Nenow}}, \binits{J.}},
\bauthor{\bsnm{{Ehgamberdiev}}, \binits{S.}},
\bauthor{\bsnm{{Khalikov}}, \binits{S.}},
\bauthor{\bsnm{{TON Team}}}:
\byear{1997},
\batitle{{Comparison of Two Fitting Methods for Ring Diagram Analysis of Very
  High L Solar Oscillations}}.
\bjtitle{\apj}
\bvolume{485},
\bfpage{869}.
doi:\doiurl{10.1086/304469}.
\end{barticle}
\endbibitem

\bibitem[\protect\citeauthoryear{{Pijpers} and {Thompson}}{1992}]{PT92}
\begin{barticle}
\bauthor{\bsnm{{Pijpers}}, \binits{F.P.}},
\bauthor{\bsnm{{Thompson}}, \binits{M.J.}}:
\byear{1992},
\batitle{{Faster formulations of the optimally localized averages method for
  helioseismic inversions}}.
\bjtitle{\aap}
\bvolume{262},
\bfpage{L33}\,--\,\blpage{L36}.
\end{barticle}
\endbibitem

\bibitem[\protect\citeauthoryear{{Pijpers} and {Thompson}}{1994}]{PT94}
\begin{barticle}
\bauthor{\bsnm{{Pijpers}}, \binits{F.P.}},
\bauthor{\bsnm{{Thompson}}, \binits{M.J.}}:
\byear{1994},
\batitle{{The SOLA method for helioseismic inversion}}.
\bjtitle{\aap}
\bvolume{281},
\bfpage{231}\,--\,\blpage{240}.
\end{barticle}
\endbibitem

\bibitem[\protect\citeauthoryear{{Rabello-Soares}}{2012}]{CRS2012}
\begin{barticle}
\bauthor{\bsnm{{Rabello-Soares}}, \binits{M.C.}}:
\byear{2012},
\batitle{{Solar-cycle Variation of Sound Speed near the Solar Surface}}.
\bjtitle{\apj}
\bvolume{745},
\bfpage{184}.
doi:\doiurl{10.1088/0004-637X/745/2/184}.
\end{barticle}
\endbibitem

\bibitem[\protect\citeauthoryear{{Rabello-Soares}, {Basu}, and
  {Christensen-Dalsgaard}}{1999}]{R-Setal99}
\begin{barticle}
\bauthor{\bsnm{{Rabello-Soares}}, \binits{M.C.}},
\bauthor{\bsnm{{Basu}}, \binits{S.}},
\bauthor{\bsnm{{Christensen-Dalsgaard}}, \binits{J.}}:
\byear{1999},
\batitle{{On the choice of parameters in solar-structure inversion}}.
\bjtitle{\mnras}
\bvolume{309},
\bfpage{35}\,--\,\blpage{47}.
doi:\doiurl{10.1046/j.1365-8711.1999.02785.x}.
\end{barticle}
\endbibitem

\bibitem[\protect\citeauthoryear{{Rajaguru}, {Basu}, and {Antia}}{2001}]{Raj01}
\begin{barticle}
\bauthor{\bsnm{{Rajaguru}}, \binits{S.P.}},
\bauthor{\bsnm{{Basu}}, \binits{S.}},
\bauthor{\bsnm{{Antia}}, \binits{H.M.}}:
\byear{2001},
\batitle{{Ring Diagram Analysis of the Characteristics of Solar Oscillation
  Modes in Active Regions}}.
\bjtitle{\apj}
\bvolume{563},
\bfpage{410}\,--\,\blpage{418}.
doi:\doiurl{10.1086/323780}.
\end{barticle}
\endbibitem

\end{thebibliography}

\end{article}
\end{document}